\documentclass[sigconf]{acmart}
\usepackage[utf8]{inputenc}

\usepackage{amsmath,amsfonts}
\usepackage{algorithmic}
\usepackage{graphicx}
\usepackage{textcomp}
\usepackage{xcolor}

\usepackage{csquotes}
\usepackage{url}
\usepackage{mathtools}
\usepackage{csquotes}
\usepackage{enumitem}
\usepackage{xspace}
\usepackage{algorithmic}
\usepackage{algorithm}
\usepackage{subcaption}
\usepackage{array}
\usepackage{comment}
\usepackage[group-separator={,}]{siunitx}
\usepackage{bm}
\usepackage{placeins}
\usepackage{amsthm}
\usepackage{xcolor}
\usepackage{soul}
\usepackage{multirow}
\usepackage{makecell}
\usepackage{wrapfig}
\usepackage{booktabs}
\usepackage{hyperref}
\usepackage{multibib}
\usepackage[normalem]{ulem}
\usepackage{todonotes}

\newcommand{\lightgcn}{\textsc{LightGCN}~}
\newcommand{\algname}{\textsc{Navip}}
\newcommand{\alexa}{Alexa~}

\title{Debiasing Neighbor Aggregation \protect\\for Graph Neural Network in Recommender Systems}
\author{Minseok Kim}
\affiliation{
    \institution{Amazon Alexa AI}
    \country{Seattle, WA, USA}
}
\email{kminseok@amazon.com}
\author{Jinoh Oh}
\affiliation{
    \institution{Amazon Alexa AI}
    \country{Seattle, WA, USA}
}
\email{ojino@amazon.com}
\author{Jaeyoung Do}
\affiliation{
    \institution{Amazon Alexa AI}
    \country{Seattle, WA, USA}
}
\email{domjae@amazon.com}
\author{Sungjin Lee}
\affiliation{
    \institution{Amazon Alexa AI}
    \country{Seattle, WA, USA}
}
\email{sungjinl@amazon.com}

\AtBeginDocument{%
  \providecommand\BibTeX{{%
    \normalfont B\kern-0.5em{\scshape i\kern-0.25em b}\kern-0.8em\TeX}}}

\setcopyright{acmcopyright}
\copyrightyear{2022}
\acmYear{2022}
\setcopyright{acmlicensed}\acmConference[CIKM '22]{Proceedings of the 31st
ACM International Conference on Information and Knowledge
Management}{October 17--21, 2022}{Atlanta, GA, USA}
\acmBooktitle{Proceedings of the 31st ACM International Conference on
Information and Knowledge Management (CIKM '22), October 17--21, 2022,
Atlanta, GA, USA}
\acmPrice{15.00}
\acmDOI{10.1145/3511808.3557576}
\acmISBN{978-1-4503-9236-5/22/10}

\acmConference[CIKM '22]{The 31st ACM International Conference on Information and Knowledge Management}{October 17--22, 2022}{Atlanta, GA, USA}

\begin{document}

\begin{abstract}
Graph neural networks (GNNs) have achieved remarkable success in recommender systems by representing users and items based on their historical interactions. However, little attention was paid to GNN's vulnerability to \emph{exposure bias}:
users are exposed to a limited number of items so that a system only learns a biased view of user preference to result in suboptimal recommendation quality.
Although inverse propensity weighting is known to recognize and alleviate exposure bias, it usually works on the final objective with the model outputs, whereas GNN can also be biased during neighbor aggregation. 
In this paper, we propose a simple but effective approach, neighbor aggregation via inverse propensity\,(\algname{}) for GNNs.
Specifically, given a user-item bipartite graph, we first derive propensity score of each user-item interaction in the graph. 
Then, inverse of the propensity score with Laplacian normalization is applied to debias neighbor aggregation from exposure bias.
We validate the effectiveness of our approach through our extensive experiments on two public and Amazon Alexa datasets where the performance enhances up to $14.2\%$.\looseness=-1

\end{abstract}

\maketitle

\vspace{-0.2cm}
\section{Introduction}

Undoubtedly, collaborative filtering\,\cite{resnick1994grouplens, konstan1997grouplens, he2017neural} became essential for personalized recommender systems in many online services.
Regarding users and items as nodes and their interactions as edges in a user-item bipartite graph, a typical collaborative filtering represents nodes into latent embedding and trains them to encode the characteristics of users and items.
In other words, this approach neglects connections among nodes that provide rich information about similarity among users and items.
In this regard, recent studies exploited graph neural networks\,(GNNs) to complement embedding by aggregating information of interacted nodes\,(i.e., neighbors) and obtained meaningful performance gain for recommender systems\,\cite{ying2018graph, wang2019neural, he2020lightgcn}.

\begin{figure}[t!]
\begin{center}
\includegraphics[width=0.45\textwidth]{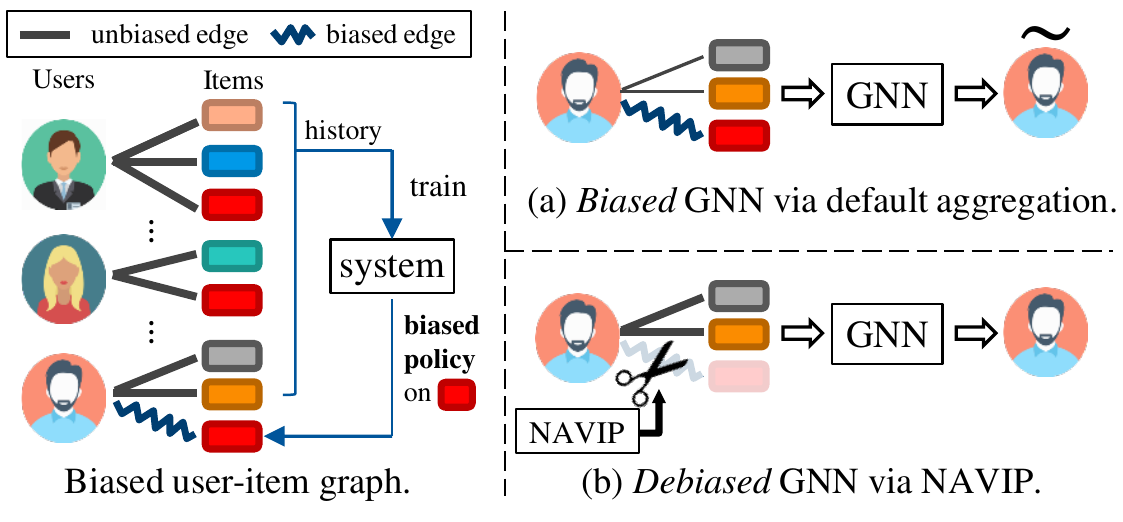}
\end{center}
\vspace*{-0.6cm}
\caption{Comparison of (a) \emph{biased} default GNN and (b) \emph{debiased} GNN via \algname{} given biased user-item graph.}
\label{fig:motivating_example}
\vspace*{-0.6cm}
\end{figure}

However, as shown in Figure \ref{fig:motivating_example}(a), real-world user-item graphs are often biased because edges are user responses to the system behavior decided by a policy. 
Specifically, a system can expose only some part of all items to users so that how user-item interactions occur becomes biased by the system's policy.
Since GNNs represent a node by aggregating neighbors connected via edges, GNNs induce a biased view of user preference, which may \emph{not} be aligned with true user interest.
As such, recommender systems based on GNNs leave room for improvement through debiasing the neighbor aggregation process with respect to the exposure bias.

A common way to mitigate exposure bias includes inverse propensity scoring\,(IPS) that emphasizes less-exposed items during training\,\cite{swaminathan2015self, schnabel2016recommendations, wang2016learning, joachims2017unbiased, gruson2019offline, saito2019unbiased, damak2021debiased}. 
Specifically, IPS multiplies the inverse propensity of user-item interactions in the objective function, which accordingly upweights the learning on less-exposed items.
However, since IPS deals with the bias problem \emph{after} GNNs encode users and items into embeddings, the system is still susceptible to the bias caused by the unadjusted process of neighbor aggregation.

In this paper, we propose a simple yet effective approach, neighbor aggregation via inverse propensity\,(\algname{}) for GNNs.
Specifically, given a user-item bipartite graph, we first derive the propensity score of each user-item interaction in the graph. 
We then use the inverse propensity score with Laplacian normalization as edge weight for the neighbor aggregation process.
As such, less popular neighbors are relatively emphasized in an embedding to balance out the biased local structure of each target node as shown in Figure \ref{fig:motivating_example}(b).
We validate the effectiveness of \algname{} with a comprehensive experiment on three real-world datasets including a large-scale voice assistant system, Amazon Alexa.
\vspace{-0.1cm}

\vspace{-0.15cm}
\section{Preliminary and Related Work}
In online services, user-item interactions data $\{(u,i,Y_{u,i})\}\in \mathcal{D}_{\pi_0}$ is collected according to \emph{policy $\pi_0$}, which is a prior on user behaviors.
\emph{Recommender systems} perform personalized recommendation by finding similar users and items based on the collected data $\mathcal{D}_{\pi_0}$, which is known as collaborative filtering\,\cite{resnick1994grouplens, konstan1997grouplens, he2017neural}.
Specifically, given a recommender model $\phi(\cdot)$ and its prediction for user $u$'s relevance on item $i$ as $\phi(h_u,h_i)$ based on their embeddings, $h_u$ and $h_i$, we minimize empirical risk $R(\pi_0)$ on implicit feedback $Y_{u,i}$,
\begin{equation}
    R(\pi_0) =\!\!\!\!\!\!\!\!\sum_{(u,i, Y_{u,i}) \in \mathcal{D}_{\pi_0}}\!\!\!\!\!\!\!\!\mathcal{L}\big(\phi(h_u,h_i), Y_{u,i}\big),
\label{eq:default_loss}
\end{equation}
where $\mathcal{L}$ is a loss function such as BPR\,\cite{rendle2012bpr}, and $Y_{u,i}$ is $1$ if the interaction between user $u$ and item $i$ is observed or $0$ if not.

A prevalent approach for collaborative filtering follows two steps: \emph{(i)} transforms each user and item into a latent representation, $h_u$ and $h_i$, then \emph{(ii)} exploits the representations to estimate the relevance $\phi(h_u,h_i)$ based on dot product\,\cite{rendle2012bpr} or non-linear neural networks\,(DNN)\,\cite{he2017neural} based on $h_u$ and $h_i$.

\subsection{GNN for Recommender System}
\label{sec:related_work_gnn}
GNN is the state-of-the-art approach for learning representation over graphs\,\cite{he2020lightgcn}.
The biggest difference between GNN and conventional CF methods is that GNN explicitly leverages neighbors of a target node during inference of the target node's representation.
Specifically, given a bipartite graph of users and items as nodes, the embedding $h$ of a user $u$ at layer $l$ is computed as follows,
\begin{align}
h^{l}_{u} &= {\rm TRANSFORM}\big(h^{l-1}_{u}, a^{l}_{u} \big)
\label{eq:transformation}\\
a^{l}_{u} &= {\rm AGGREGATE}\big(h^{l-1}_{i}, i \in \mathcal{N}(u) \big),
\label{eq:aggregation}
\end{align}
where $a^{l}_{u}$ denotes the aggregated embedding of $u$'s neighbors ${\mathcal{N}(u)}$ at layer $l$. 
Note that we can augment the embedding of item $i$ in the same way. 
The design of ${\rm AGGREGATE(\cdot)}$ and ${\rm TRANSFORM(\cdot)}$, or leveraging high-order connectivity are used to distinguish GNN methods.
Initial GNN methods\,\cite{kabbur2013fism, koren2008factorization, berg2017graph} used only one-hop neighbors using one GNN layer, while following studies\,\cite{ying2018graph, wang2019neural} exploit multiple layers with non-linear DNN to benefit from high-order connectivity.
%
Recently, \lightgcn\cite{he2020lightgcn} claimed that such non-linear DNN is not essential for GNN, and showed a state-of-the-art performance by employing symmetric Laplacian norm in ${\rm AGGREGATE(\cdot)}$ and identity function for ${\rm TRANSFORM(\cdot)}$,
\begin{equation}
h^{l}_{u} = a^{l}_{u} = \sum_{i \in \mathcal{N}(u)} \frac{h^{l-1}_{i}}{\sqrt{|\mathcal{N}(u)||\mathcal{N}(i)|}}.
\end{equation}

\subsection{Inverse Propensity Scoring}
In a service, users are exposed to only a part of items selected by a system policy $\pi_0$.
Thus, collected log data $\mathcal{D}_{\pi_0}$ becomes dependent toward the policy $\pi_0$, which is called \emph{exposure bias}.
Naturally, a recommender $\phi$ trained on $\mathcal{D}_{\pi_0}$ becomes biased and even intensifies the bias\,\cite{yang2018unbiased, zhang2021causal}.
To this end, IPS is popularly exploited to acquire unbiased estimator $R(\pi)$ of the performance on the uniform item exposure policy $\pi$ by reweighting $\mathcal{D}_{\pi_0}$ with propensity\,\cite{saito2019unbiased, saito2020unbiased},
\begin{equation}
    \hat{R}(\pi; \mathcal{D}_{\pi_0})=\frac{1}{|\mathcal{D}_{\pi_0}|}\sum_{(u,i, Y_{u,i})\in \mathcal{D}_{\pi_0}}\!\!\frac{\mathcal{L}\big(\phi(h_u,h_i), Y_{u,i}\big)}{p(Y_{u,i}\!=\!1|\pi_0)},
\label{eq:IPS_loss}
\end{equation}
where $p(Y_{u,i}\!=\!1|\pi_0)$ is propensity defined in Definition \ref{def:propensity}.
\begin{definition}
\label{def:propensity}
\emph{Propensity $p(Y_{u,i}\!=\!1|\pi_0)$} is the observation probability of user-item interaction $Y_{u,i}$ under policy $\pi_0$.
\end{definition}
\noindent 
However, IPS has a few drawbacks, which are complemented by several studies.
For example, SNIPS\,\cite{swaminathan2015self} was suggested to address high variance of IPS at the expense of a small estimation bias\,\cite{hesterberg1995weighted, schnabel2016recommendations}.
Recently, UEBPR\,\cite{damak2021debiased} improves IPS-based recommender learning by reflecting recommendation explainability during training.
However, we note that the above-described IPS studies work in the objective function \emph{after} GNN neighbor aggregation, therefore the system is sill biased even with IPS.

\section{Methodology}

Previous methods have mainly focused on applying IPS on loss as shown in Eq.\,\eqref{eq:IPS_loss}, but gave less attention to how the neighbors are aggregated even though it is one of the key factors that contribute to the success of GNN.
Because collected log data $\mathcal{D}_{\pi_0}$ has exposure bias\,(i.e., $\mathcal{D}_{\pi_0} \neq \mathcal{D}_{\pi}$), GNNs also employ neighbor aggregation to biased interactions $\{x=(u,i,Y_{u,i})~|~x \in \mathcal{D}_{\pi_0} \wedge x \notin \mathcal{D}_{\pi}\}$, which can bias user embedding $h_{u}^{l}$.
Even worse, the bias can propagate to users without biased interactions due to the repeated GNN layers.

\subsection{\algname{} for Debiasing GNN}
Inspired by the motivation, we suggest \algname{} that applies IPS weighting scheme to neighbor aggregation to reduce the impact of biased interactions and debias GNN neighbor aggregation.
That is, \algname{} can be considered as upweighting interactions with low propensity and downweighting interactions with high propensity during aggregation.
The rationale behind \algname{} is that neighbors with lower propensity interaction reveal more about the true interest of a given target user because such interactions are not likely to occur by a biased system policy.
On the other hand, because interactions with high propensity often occur as a response to a biased system policy, they may not be the true user interest.

For simplicity, we tentatively assume a simple neighbor aggregation, $a^{l}_u=\mathbb{E}[h^{l-1}_i]$, which is the expected embedding of neighbors at layer $l-1$.
For clarity, we further define $a^{l}_u(\pi)$ to be the aggregated neighbor embedding of user $u$ under the system policy $\pi$, and $a^{l}_u(\pi|\mathcal{D}_{\pi})$ to be the \emph{Monte-Carlo approximated} aggregated neighbor embedding of $\pi$ over $\mathcal{D}_{\pi}$.
Formally speaking,
\begin{equation}
\begin{aligned}
a^{l}_u(\pi) 
&\approx \frac{1}{|\mathcal{N}_{\pi}(u)|}\sum_{i\in\mathcal{N}_{\pi}(u)} h^{l-1}_i(\pi) = a^{l}_u(\pi|\mathcal{D}_{\pi})
\end{aligned}
\label{eqn:neighbor_monte}
\end{equation}
where $h^{l-1}_i(\pi)$ is a node embedding at $l-1$ layer under policy $\pi$.
In practice, $a^{l}_u(\pi)$ needs to be approximated based on $\mathcal{D}_{\pi_0}$ because $\mathcal{D}_{\pi}$ is not available while developing $\pi$.
\algname{} estimates $a^{l}_u(\pi)$ with respect to $\mathcal{D}_{\pi}$ by combining Eq.~\eqref{eq:IPS_loss} and Eq.~\eqref{eqn:neighbor_monte},
\begin{equation}
\begin{aligned}
a^{l}_u(\pi) 
&\approx \frac{1}{|\mathcal{N}_{\pi_0}(u)|}\sum_{i\in\mathcal{N}_{\pi_0}(u)} \frac{h^{l-1}_i(\pi)}{p(Y_{u,i}\!=\!1|\pi_0)} = a^{l}_u(\pi|\mathcal{D}_{\pi_0}).
\end{aligned}
\end{equation}
However, we note that using inverse propensity results in numerical instability issue due to unnormalized magnitude of neighbor weights.
To avoid the issue, \algname{} further normalizes neighbor aggregation by replacing $\frac{1}{|\mathcal{N}_{\pi_0}(u)|}$,
\begin{equation}
\frac{1}{Z_{\pi}(u)}\sum_{i\in\mathcal{N}_{\pi_0}(u)} \frac{h^{l-1}_i(\pi)}{p(Y_{u,i}\!=\!1|\pi_0)} = a^{l}_u(\pi|\mathcal{D}_{\pi_0}),
\label{eq:NAVIP_aggregation}
\end{equation}
where $Z_{\pi_0}(u)$ is a normalizing term defined by $\sum_{i\in\mathcal{N}_{\pi_0}(u)} \frac{1}{p(Y_{u,i}=1|\pi_0)}$.
We replace Eq.\,\eqref{eq:aggregation} with Eq.\,\eqref{eq:NAVIP_aggregation} to debias neighbor aggregation.

\subsubsection{Connection to Laplacian}
The neighbor aggregation function of \algname{}~can be formulated in a matrix form. Specifically,
\begin{equation}
\begin{aligned}
\Pi_{u,i} &= 
    \begin{cases}
        \frac{1}{P(Y_{u,i}=1|{\pi_0})} & \text{if} \quad Y_{u,i}=1\\
        0 & \text{else},
    \end{cases}\\
\boldsymbol{\Lambda} &=
    \begin{bmatrix}
    \boldsymbol{0}   & \boldsymbol{\Pi} \\
    \boldsymbol{\Pi}^\top     & \boldsymbol{0}.
    \end{bmatrix}
\end{aligned}
\end{equation}
Then, the weighted random walk Laplacian can be shown as
\begin{equation}
\boldsymbol{A}^{l}(\pi_0|\mathcal{D}_{\pi}) = \boldsymbol{D}^{-1}\boldsymbol{\Lambda}\boldsymbol{H}^{l}(\pi_0),
\end{equation}
where $\boldsymbol{A}^{l}$ and $\boldsymbol{H}^{l}$ are the matrix of aggregated neighbor embedding and node embedding at layer $l$, $\boldsymbol{\Lambda}$ is the adjacency matrix, and $\boldsymbol{D}$ is a diagonal degree matrix whose $k$-th diagonal elements is the sum of $k$-th row of $\boldsymbol{\Lambda}$, $D_{kk}=\sum_i \Lambda_{ki}$.

It is worth noting that $\boldsymbol{D}^{-1}\boldsymbol{\Lambda}$ can be interpreted as random walk Laplacian, $\boldsymbol{L}^{rw}$, on a \emph{weighted graph}, whose weights are inverse propensity weights.
This simple difference from default aggregation encourages neighbors with low propensity edge become dominant in ${h}^{l}_{u}$, even with more GNN layers.


\vspace{-0.2cm}
\subsection{Model Training}
The simplicity of \algname{} brings two benefits during training: \emph{(i) no computational overhead} due to \algname{} during training because \algname{} debiases GNN neighbor aggregation \emph{without} any additional training parameter, and \emph{(ii) no objective adjustment} because \algname{} modifies neighbor aggregation before final model output $\phi(h_u,h_i)$ so as to be independent of learning objectives as in Eq.\eqref{eq:default_loss} or Eq.\eqref{eq:IPS_loss}.
In this regard, \algname{} can follow given recommender's training scheme while debiasing the recommender.
In addition, it is worth noting that \algname{} can achieve the debiasing effect for the pretrained model \emph{without} additional training\,(see Section \ref{sec:overall_performance} for more details).

\section{Evaluation}

We conducted experiments to answer the following questions:
\begin{itemize}[leftmargin=9pt]
    \item Does \algname{} improve overall recommender performance?
    \item How does \algname{} impact recommender's exposure bias?
    \item Does \algname{} have overlapping effect with IPS?
\end{itemize}

\subsection{Experimental Settings}

\subsubsection{Datasets}
We conducted experiments on three real-world datasets: \alexa, Yahoo\footnote{http://webscope.sandbox.yahoo.com/}, and Movielens\footnote{http://grouplens.org/datasets/movielens/}.
\begin{itemize}[leftmargin=9pt]
    \item \alexa: it contains a truncated user interaction logs from a voice assistant service, Amazon Alexa. The interactions are collected for a week in August 2021. Redundant user-item interactions such as users' daily weather checking were filtered out to align setting with other two datasets.
    \item Yahoo: it contains users' five-star song ratings from Yahoo! Music. Different from other two datasets, Yahoo contains unbiased test data which is obtained by asking $5,400$ users to rate $10$ randomly selected songs.
    To make the ratings as implicit feedback, we transformed all the ratings into $Y_{u,i} = 1$ and unobserved entries as $Y_{u,i} = 0$ to indicate whether a user $u$ interacted with an item $i$.
    \item Movielens: it contains users' five-star ratings for movies. Like Yahoo, all the ratings were binarized.
\end{itemize}

We filtered out users and items that contain less than 10 interactions following usual practice\,\cite{he2017neural}.
Table \ref{Tab:dataset} shows the profile of the three datasets after preprocessing.

\begin{table}[t]
\centering
\footnotesize
\begin{tabular}{ccccc}
\toprule
Dataset     & Users  &Items  & Interactions  \\ \midrule
\alexa                  & 40k               & 2.5k              & 0.7m \\
Yahoo               & 15k               & 1.0k              & 0.3m                \\ 
Movielens          & 943               & 1.6k              & 0.1m                \\ \bottomrule
\end{tabular}
\caption{Summary of three real-world datasets.}
\label{Tab:dataset}
\vspace{-0.8cm}
\end{table}

We created pseudo-unbiased testing and validation data for \alexa~and Movielens dataset.
Because real-world datasets are biased, test data created by random splitting is not effective to evaluate the debiasing effect.
Having a dedicated unbiased test dataset generated from pure random exposure is ideal, but this is practically infeasible in many cases. 
Following~\cite{saito2019unbiased, damak2021debiased}, we generated pseudo-unbiased test data by sampling $5\%$ of total data based on the inverse of relative item popularity.
In case of Yahoo, we used the original test data, but for validation set, we followed the same process to generate a unbiased validation data.


\subsubsection{Implementation detail.}
We used \lightgcn\cite{he2020lightgcn}\footnote{https://github.com/gusye1234/LightGCN-PyTorch}, which is the state-of-the-art GNN-based recommender system, as a backbone GNN throughout the experiment.
We investigated three neighbor aggregations to validate the effect of neighbor aggregation.
\begin{itemize}[leftmargin=9pt]
\item \textsc{Mean}: a basic neighbor aggregation strategy that averages embeddings of neighbors.
\item \textsc{Propensity}: a bias-oriented neighbor aggregation strategy with propensity $P(Y_{u,i}=1|\pi_0)$ as edge weight. 
\item \algname{}: a \emph{debiasing} neighbor aggregation strategy that uses IPS as edge weight for each neighbor as in Eq.\eqref{eq:NAVIP_aggregation}. 
\end{itemize}
In addition, to validate whether the effect of \algname{} overlaps with that of IPS methods, we conducted experiment with the state-of-the-art IPS method for ranking loss, \textsc{UEBPR}\,\cite{damak2021debiased}.

Following~\cite{saito2019unbiased,damak2021debiased}, we estimate the propensity of an item to a user by relative item popularity,
\begin{equation}
    p(Y_{u,i}\!=\!1|\pi_0)=\sqrt{\frac{\sum_{u\in U} Y_{u,i}}{\text{max}_{l\in I} \sum_{u\in U} Y_{u,l}}},
    \label{eq:relative_item_popularity}
\end{equation}
where $U$ and $I$ are the set of users and items, respectively.

All the experiments were trained for $100$ epochs with batch size $256$.
The size of input feature and embedding are $64$.
We used Adam\,\cite{kingma2014adam} with a learning rate $\eta = 0.003$.
Our implementation was written using Pytorch and tested on Nvidia Tesla V100.
We conducted experiment $10$ times and reported their average.

\subsubsection{Evaluation Metrics}
We used two widely-recognized metrics, hit rate\,(HR) and normalized discounted cumulative gain\,(NDCG).
Given a model's recommendation list, HR measures the proportion of true user-item interactions in a recommended list, and NDCG additionally reflects the ranking of interactions.
Two metrics were calculated with top-$k$ recommendation: a model chooses $k$ most probable items that each user is expected to like and suggest them to the user.
We varied the number of recommended items $k\in\{5, 10, 20\}$.
Because it is time-consuming to rank all items, a recommender evaluated the rank of the user's $1$ interacted item among randomly sampled $99$ non-interacted items following literature\,\cite{he2016fast}.

\subsection{Overall Performance}
\label{sec:overall_performance}
Table \ref{Tab:performance_only_in_test} shows the performance of pretrained \lightgcn~model with varying neighbor aggregation during inference to evaluate the effect of neighbor aggregation.
The performance difference among neighbor aggregation methods verifies that neighbor aggregation \emph{significantly} affects the quality of embedding so as to impact on the success of recommendation.
Overall, \algname{}~shows the best performance among three neighbor aggregations in most cases.

Specifically, \algname{} outperforms \textsc{Mean} and \textsc{Propensity} by up to $14.2\%$ and $126\%$ in terms of $NDCG@5$ in \alexa~dataset, respectively.
Such performance improvement is attributed to the debiasing effect of \algname{}~on neighbor aggregation, which successfully downweights neighbors with biased interaction so as to receive relatively more information from relevant neighbors.   
This claim can be further supported by the performance of \textsc{Propensity} which always recorded the worst performance among three neighbor aggregations.
That is, \textsc{Propensity} highlights neighbors with biased interactions, which may not represent the true interest of users.\looseness=-1

We also investigate the impact of neighbor aggregation when they are used in both training and inference (Table \ref{Tab:performance_train_and_test}).
The performance gap among neighbor aggregation strategies decreased so that all three aggregations showed performance similar to \textsc{Mean}.
We conjecture that training learnable \lightgcn input features under the same learning objective can weaken the effect of different neighbor aggregation.
Nevertheless, the order of recommendation performance among neighbor aggregation strategies maintains; \algname{} scored the best performance in general, while \textsc{Propensity} consistently performs the worst.

\begin{table}
\begin{center}
\resizebox{\linewidth}{!}{
\begin{tabular}{c|c|ccc|ccc}\toprule
\centering{\multirow{2}{*}{Dataset}} & \centering{\multirow{2}{*}{Method}}
& \multicolumn{3}{c}{HR}   \vline & \multicolumn{3}{c}{NDCG} \\ \cline{3-8}
\rule{0pt}{2.5ex} & & @5 & @10 & @20 & @5 & @10 & @20 \\[-0.25ex] \midrule
\multirow{3}{*}{\alexa}     & \textsc{Mean}        & 0.266          & 0.364          & 0.498          & 0.190          & 0.222          & 0.256          \\
                          & \textsc{Propensity}  & 0.153          & 0.255          & 0.410          & 0.096          & 0.129          & 0.167          \\
                          & \algname{}       & \textbf{0.287} & \textbf{0.379} & \textbf{0.509} & \textbf{0.217} & \textbf{0.247} & \textbf{0.279} \\ \midrule
\multirow{3}{*}{Yahoo}     & \textsc{Mean}        & 0.145          & 0.242          & 0.397          & 0.093          & 0.124          & 0.163          \\
                          & \textsc{Propensity}  & 0.134          & 0.224          & 0.376          & 0.086          & 0.115          & 0.153          \\
                          & \algname{}       & \textbf{0.148} & \textbf{0.249} & \textbf{0.408} & \textbf{0.095} & \textbf{0.127} & \textbf{0.167} \\ \midrule
\multirow{3}{*}{Movielens} & \textsc{Mean}        & \textbf{0.350} & 0.506          & 0.681          & \textbf{0.233} & \textbf{0.283} & 0.327          \\
                          & \textsc{Propensity}  & 0.343          & 0.493          & 0.668          & 0.230          & 0.278          & 0.322          \\
                          & \algname{}       & 0.349 & \textbf{0.511} & \textbf{0.691} & 0.230 & \textbf{0.283} & \textbf{0.328} \\ \bottomrule
\end{tabular}
}
\end{center}
\caption{Recommendation performance of \emph{pretrained} \lightgcn varying neighbor aggregation method \emph{during testing.}}
\label{Tab:performance_only_in_test}
\vspace*{-0.8cm}
\end{table}

\begin{table}
\resizebox{\linewidth}{!}{
\begin{tabular}{c|c|ccc|ccc}\toprule
\centering{\multirow{2}{*}{Dataset}} & \centering{\multirow{2}{*}{Method}}
& \multicolumn{3}{c}{HR}   \vline & \multicolumn{3}{c}{NDCG} \\ \cline{3-8}
\rule{0pt}{2.5ex} & & @5 & @10 & @20 & @5 & @10 & @20 \\[-0.25ex] \midrule
\multirow{3}{*}{\alexa}
& \textsc{Mean}
 & 0.266          & 0.364          & 0.498          & 0.190          & 0.222          & 0.256        
\\
& \textsc{Propensity}
& 0.260          & 0.357          & 0.489          & 0.187          & 0.218          & 0.251  
\\
& \algname{}
& \textbf{0.268} & \textbf{0.368} & \textbf{0.504} & \textbf{0.192} & \textbf{0.224} & \textbf{0.258} 
\\
\midrule
\multirow{3}{*}{Yahoo}
& \textsc{Mean}
& 0.145          & 0.242          & \textbf{0.397} & 0.093          & 0.124          & 0.163        
\\
& \textsc{Propensity}
& 0.145          & 0.241          & 0.392          & 0.093          & 0.124          & 0.162     
\\
& \algname{}
& \textbf{0.147} & \textbf{0.246} & \textbf{0.397} & \textbf{0.095} & \textbf{0.126} & \textbf{0.164} 
\\
\midrule
\multirow{3}{*}{Movielens}
& \textsc{Mean}
& 0.350 & 0.506          & 0.681          & 0.233 & 0.283 & 0.327  
\\
& \textsc{Propensity}
& 0.347          & 0.503          & 0.678          & 0.233          & 0.283          & 0.327  
\\
& \algname{}
& \textbf{0.352} & \textbf{0.512} & \textbf{0.687} & \textbf{0.234} & \textbf{0.286} & \textbf{0.330}
\\
\bottomrule
\end{tabular}
}
\caption{Recommendation performance of each neighbor aggregation method \emph{during training and testing.}}
\label{Tab:performance_train_and_test}
\vspace{-0.8cm}
\end{table}

\subsection{Impact on Recommender's Exposure Bias}
Table \ref{Tab:Itemwise_recommendation_performance} shows the performance of \lightgcn on head items with top $10\%$ popularity and tail items with bottom $10\%$ popularity.
The result shows that the performance on head items is higher than on tail items for all datasets, because recommenders are known to expose popular items more to users\,\cite{abdollahpouri2019unfairness, mansoury2020feedback}.
\algname{} alleviates such issues by emphasizing user interest in tail items, which leads to more exposure to tail items.
That is, \algname{} encourages \lightgcn to aggregate more information from neighbors with a low propensity, which correspond to tail items, so that \lightgcn can reveal user interest in tail items.
Although \algname{} may lead to a performance drop on head items as in Yahoo and Movielens dataset, it is negligible because \algname{} improves overall performance.

\begin{table}
\resizebox{\linewidth}{!}{
\begin{tabular}{c|c|ccc|ccc}\toprule
\multirow{2}{*}{Data}          & \multirow{2}{*}{Method} & \multicolumn{3}{c|}{Head Items}         & \multicolumn{3}{c}{Tail Items}            \\ \cline{3-8} 
                               &                              &\rule{0pt}{2.5ex} HR@5           & HR@10          & HR@20          & HR@5           & HR@10          & HR@20          \\\midrule
\multirow{2}{*}{\alexa}         & Mean                         & 0.268          & 0.369          & 0.505          & 0.264          & 0.361          & 0.494          \\
                               & \algname{}                        & \textbf{0.271} & \textbf{0.373} & \textbf{0.510} & \textbf{0.265} & \textbf{0.363} & \textbf{0.499} \\\midrule
\multirow{2}{*}{Yahoo}         & Mean                         & \textbf{0.497} & \textbf{0.638} & \textbf{0.795} & 0.033          & 0.071          & 0.141          \\
                               & \algname{}                        & 0.483          & 0.631          & 0.781          & \textbf{0.038} & \textbf{0.076} & \textbf{0.159} \\\midrule
\multirow{2}{*}{Movielens} & Mean                         & \textbf{0.649}          & \textbf{0.812}          & \textbf{0.933}          & 0.077          & 0.142          & 0.236          \\
                               & \algname{}                        & 0.647          & 0.811          & 0.931          & \textbf{0.080} & \textbf{0.149} & \textbf{0.250} \\ \bottomrule
\end{tabular}
}
\caption{Accuracy on head\,(top $10\%$ popularity) and tail\,(bottom $10\%$ popularity) items with respect to two neighbor aggregations, \textsc{Mean} and \algname{}.}
\label{Tab:Itemwise_recommendation_performance}
\vspace*{-0.5cm}
\end{table}

\subsection{Relationship with IPS}
Table \ref{Tab:performance_with_IPS} shows the performance of \lightgcn trained with IPS method \textsc{UEBPR} during training.
While \algname{} with IPS showed the best performance in \alexa\,and Yahoo datasets for both types of Laplacian norm, \algname{} without IPS scored the best in Movielens dataset.
We point out that such inconsistent results in Movielens arose from the high variance of IPS that may lead training to be far from the ideal training objective $R(\pi)$\,\cite{swaminathan2015self}.
Nevertheless, we can conclude that \algname{} and IPS can be synergetic in debiasing GNN.

\begin{table}
\resizebox{\linewidth}{!}{
\begin{tabular}{c|c|ccc|ccc}\toprule
\centering{\multirow{2}{*}{Dataset}} & \centering{\multirow{2}{*}{Method}}
& \multicolumn{3}{c}{HR}   \vline & \multicolumn{3}{c}{NDCG} \\ \cline{3-8}
\rule{0pt}{2.5ex} & & @5 & @10 & @20 & @5 & @10 & @20 \\[-0.25ex] \midrule
\multirow{2}{*}{\alexa}
& \textsc{Mean}
&  0.299    &	0.400 &	0.533   &   0.218   &	0.251   &	0.284    
\\
& \algname{}
& \textbf{0.301}  &	\textbf{0.403}  & \textbf{0.538} &   \textbf{0.220} &	\textbf{0.253}  &	\textbf{0.287}	
\\
\midrule
\multirow{2}{*}{Yahoo}
& \textsc{Mean}
& 0.152          & 0.250          & 0.408         & \textbf{0.098} & \textbf{0.130}  & 0.169 
\\
& \algname{}
& \textbf{0.153} & \textbf{0.252} & \textbf{0.411} & \textbf{0.098} & \textbf{0.130} & \textbf{0.170} 
\\

\midrule
\multirow{2}{*}{Movielens}
& \textsc{Mean} 
& 0.350          & \textbf{0.512} & \textbf{0.699} & 0.230          & 0.282          & \textbf{0.330}
\\
& \algname{}
& 0.349          & \textbf{0.512} & 0.693         & 0.229           & 0.282          & 0.328       

\\

\bottomrule
\end{tabular}
}
\caption{Recommendation performance with \textsc{UEBPR} on two neighbor aggregation methods, \textsc{Mean} and \algname{}. The results that exceeds the performance in Table \ref{Tab:performance_train_and_test} are marked in bold.}
\vspace*{-0.8cm}
\label{Tab:performance_with_IPS}
\end{table}

\section{Conclusion}
In this paper, we proposed \algname{} to debias in exposure bias during neighbor aggregation of GNN.
\algname{} acquires inverse propensity score for each user-item interaction and exploits its Laplacian norm as neighbor weight.
Experiments conducted on three real-world datasets confirmed that \algname{}~can successfully debias GNN neighbor aggregation and enhance performance up to $14.2\%$.

\bibliographystyle{ACM-Reference-Format}
\bibliography{reference}

\end{document}